\documentclass[aps,prl,twocolumn,superscriptaddress,showpacs,amsmath,amssymb,floatfix]{revtex4}
\usepackage{graphicx}% Include figure files
\usepackage{dcolumn}% Align table columns on decimal point
\usepackage{bm}% bold math

\begin{document}

%\preprint{ICM2003:4-ypm20}

\title{Spin Fluctuation in Heavy Fermion 
CeRu$_2$Si$_2$}

\author{Hiroaki Kadowaki}
%\email{kadowaki@phys.metro-u.ac.jp}
\affiliation{Department of Physics, Tokyo Metropolitan University, 
Hachioji-shi, Tokyo 192-0397, Japan}

\author{Masugu Sato}
\affiliation{MSD, JASRI, 
1-1-1 Kouto Mikazuki-cho Sayo-gun, Hyogo 679-5198, Japan}

\author{Shuzo Kawarazaki}
\affiliation{Department of Earth and Space Science, Osaka University, 
Toyonaka, Osaka 560-0043, Japan}

\date{\today}

\begin{abstract}
Spin fluctuations of the archetypal heavy-fermion compound CeRu$_2$Si$_2$ 
have been investigated by neutron scattering in an entire irreducible 
Brillouin zone. The dynamical susceptibility is remarkably well described 
by the self-consistent renormalization (SCR) theory of the spin fluctuation 
in a phenomenological way, proving the effectiveness of the theory. The 
present analysis using the SCR phenomenology has allowed us to determine 
fourteen exchange constants, which show the long-range nature of the 
Ruderman-Kittel-Kasuya-Yosida interaction.
% ( \sim 549 char <= 600)
%
\end{abstract}

\pacs{
75.30.Mb, 71.27.+a, 75.30.Et
}

%\keywords{ Suggested keywords }

\maketitle

Effects of the strong correlation of $d$- and $f$-electron systems are 
exhibited in dual aspects of localized and itinerant characters 
\cite{MoriyaBook85,Moriya-Ueda00,Tsunetsugu97}. 
In heavy-fermion systems, observations with energies larger than a 
small scale, Kondo temperature $T_{\text{K}}$, show local-moment 
behavior, such as the Curie-Weiss susceptibility, and antiferromagnetic 
correlations \cite{Tsunetsugu97}. 
While at lower energy scales $f$ and conduction electrons form composite 
quasiparticles with a large mass renormalization $m^*/m \propto C/T$ by a 
factor of up to a few thousands. 
This large effective mass has been ascribed to the local Kondo effect 
and to nearness to a quantum critical point (QCP) at $T=0$, 
which separates the heavy Fermi-liquid (FL) state from an antiferromagnetic 
phase of local moments interacting with Ruderman-Kittel-Kasuya-Yosida (RKKY) 
exchange interactions. 
Theoretical treatment of the both localized and itinerant characters has 
been a difficult and central issue for heavy fermions \cite{Kuramoto90}. 
In fact, a focus of recent experimental \cite{Stewart01} and theoretical 
\cite{Coleman01} studies on systems close to QCP, or non-Fermi-liquid 
behavior, is directed toward revealing the nature of the fixed point, i.e., 
whether it is a spin-density-wave type 
\cite{Lohneysen01,Moriya-Taki,Hertz76-Millis93} 
or a locally critical quantum phase transition 
\cite{Schroder00,Coleman01,Si01}.

Attempts to dealing with spin fluctuations in heavy fermions including 
the both localized and itinerant characters in a wide region close to 
QCP have been carried out by the self-consistent renormalization (SCR) 
theory of the spin fluctuation \cite{Moriya-Taki}. 
This extends the SCR theory, well established for weak ferromagnets and 
antiferromagnets of $d$-electron systems 
\cite{MoriyaBook85,Moriya-Ueda00}, 
to $f$-electron heavy-fermions using the same form of the dynamical 
susceptibility $\chi (\bm{q}, E)$ in the phenomenological way 
\cite{Moriya-Taki,Moriya-Ueda00}. 
Experimental results of bulk properties, such as temperature dependence 
of $C/T$, have been successfully analyzed using the SCR theory for many 
heavy fermions especially for paramagnetic states \cite{Moriya-Taki,Kambe97}. 
However, an experimental test \cite{Kambe96,Raymond97} of the SCR theory 
using the archetypal heavy-fermion CeRu$_2$Si$_2$ \cite{Besnus85,Haen87} 
by means of neutron scattering and bulk quantities showed that measured 
$\chi (\bm{q}, E)$ is only semi-quantitatively consistent with $C/T$, 
posing a question about the phenomenology. 
Extending this work, the present study is aimed at performing a more 
rigorous test of the SCR theory on CeRu$_2$Si$_2$ by comprehensive 
measurements of neutron scattering. 
We have shown that the SCR theory describes $\chi (\bm{q}, E)$ remarkably 
well, and that it also provides a rewarding method to determine a number 
of exchange constants. 

\begin{figure}
\begin{center}
\includegraphics[width=8.7cm,clip]{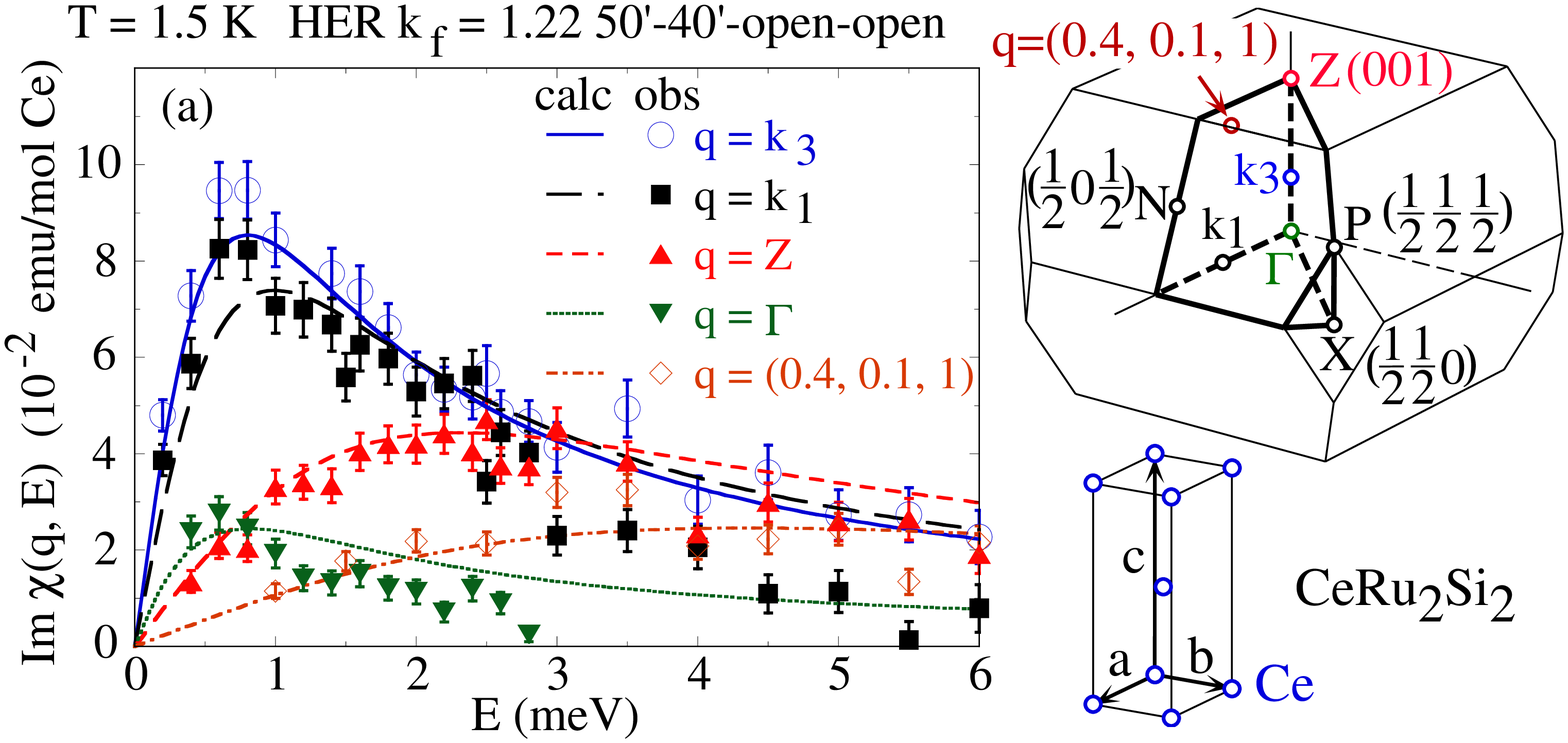}
\includegraphics[width=8.7cm,clip]{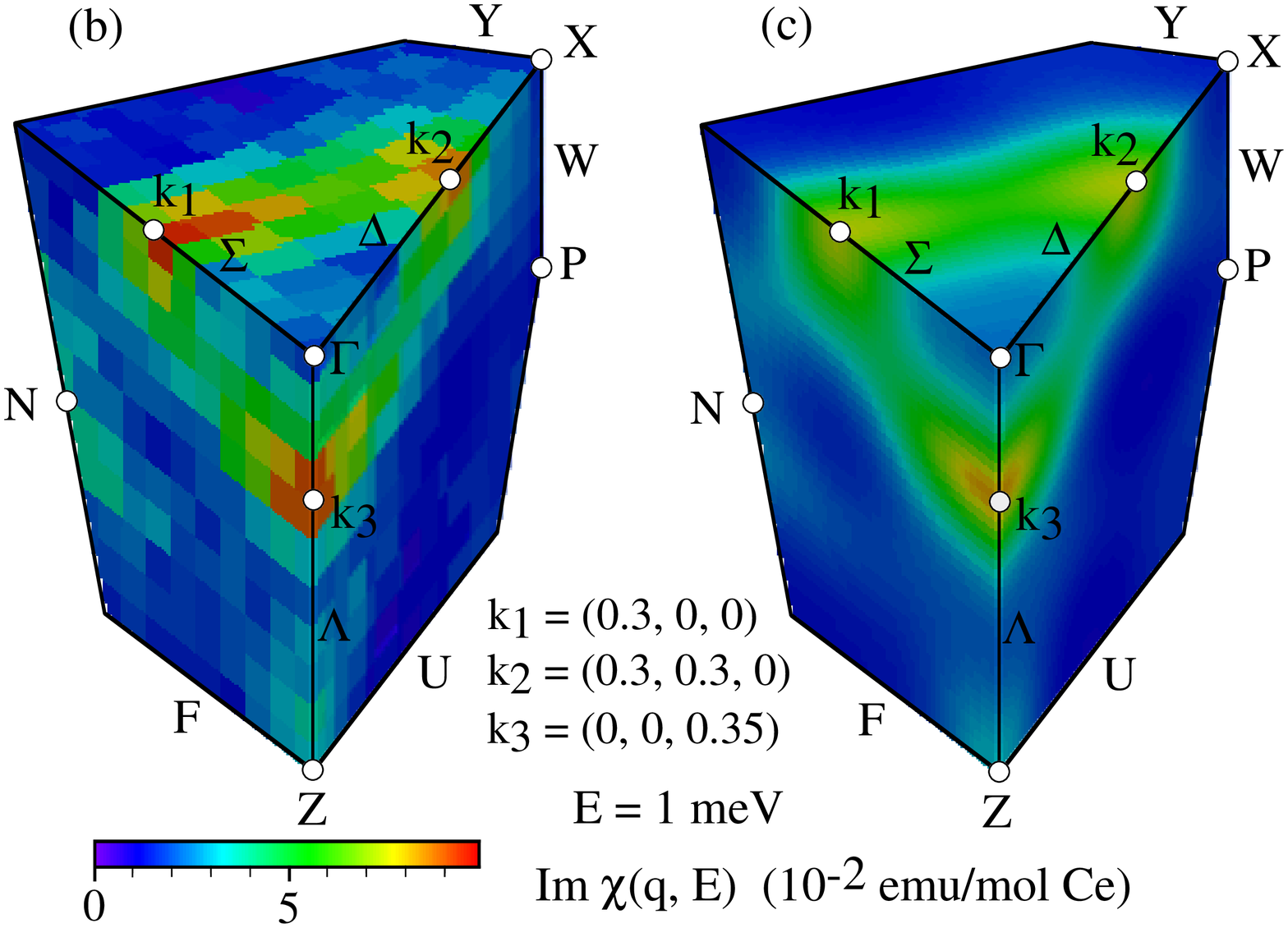}
\end{center}
\caption{
\label{fig:3Dmap}
(a) Constant-$Q$ scans at several wave vectors reduced to the 
irreducible Brillouin zone, which is illustrated on the right side 
together with the body-centered tetragonal structure. 
Curves are calculated by the fit to the SCR theory, i.e., 
Eqs.~(\ref{eq:ImChi}). 
(b) An intensity map of constant-$E$ scans taken with $E = 1$ meV 
at $T=1.5$ K is shown on the surface of the irreducible Brillouin zone. 
Antiferromagnetic spin correlations show three peaks at wave vectors 
$\bm{k}_1 = (0.3, 0, 0)$, 
$\bm{k}_2 = (0.3, 0.3, 0)$, 
and $\bm{k}_3 = (0, 0, 0.35)$. 
(c) Calculated intensity map using the fit to the SCR theory. 
}
\end{figure}

CeRu$_2$Si$_2$, which crystallizes in a body-centered tetragonal structure 
[the ThCr$_2$Si$_2$ structure, see Fig.~\ref{fig:3Dmap}(a)], is an 
archetypal heavy-fermion compound with enhanced 
$C/T \simeq 350$ mJ/K$^2$ mol \cite{Besnus85,Haen87}. 
It shows Kondo behavior with $T_{\text{K}} \simeq 24$ K \cite{Besnus85} 
and remains in a paramagnetic state down to the lowest temperature 
investigated. 
The local moments with strongly uniaxial anisotropy develop antiferromagnetic 
correlations with the energy scale of $k_{\text{B}} T_{\text{K}}$ 
\cite{Regnault88,JRM88}. 
For $T \ll T_{\text{K}}$ CeRu$_2$Si$_2$ exhibits FL behavior, for example 
$\rho - \rho_0 \propto T^2$ \cite{Haen87}, with renormalized heavy 
quasiparticles, proved by the de Haas-van Alphen (dHvA) effect 
%\cite{Lonzarich88,Zwicknagl92,Aoki93}. 
\cite{Lonzarich88-Aoki93,Zwicknagl92}. 
An intriguing property of this compound is the metamagnetic behavior under 
a magnetic field $H_{\text{M}} = 7.7$ T, showing a sharp crossover of 
magnetic states 
%\cite{Haen87,Aoki93,Raymond98,Sato-Ohkawa01}. 
\cite{Haen87,Lonzarich88-Aoki93,Raymond98,Sato-Ohkawa01}. 

Neutron-scattering measurements were performed on the triple-axis 
spectrometers HER and GPTAS at JAERI.
%, equipped with PG(002) monochromator and analyzer. 
Typical energy resolutions using final energies of
$E_{\text{f}}=3.1$ and $13.7$ meV were $0.1$ and $1.0$ meV 
(full width at half maximum), respectively, at elastic positions. 
Four single crystals with a total weight of 19 g were grown by the 
Czochralski method. 
They were aligned together and mounted in a He flow cryostat. 
All the data shown are converted to the dynamical susceptibility 
and corrected for the magnetic form-factor and the orientation factor. 
It is scaled to absolute units by comparison with the intensity of 
%the incoherent scattering from 
a standard vanadium sample. 

The dynamical susceptibility $\chi (\bm{q}, E)$ is assumed, in the 
SCR theory \cite{Moriya-Taki}, to be described by 
\begin{equation}
\label{eq:DS}
\chi (\bm{q}, E)^{-1} 
= 
\chi_{\text{L}} (E)^{-1} 
- J(\bm{q})
\; ,
\end{equation}
where the local dynamical susceptibility 
$\chi_{\text{L}} (E) = 
\chi_{\text{L}}/(1 - \text{i} E/ \Gamma_{\text{L}})$, 
expressing the local quantum fluctuation by the Kondo effect, 
is modulated by the intersite exchange interactions 
$J_{\bm{r}, \bm{r}'}$, and $J(\bm{q}) = 
\sum_{\bm{r} \neq 0} J_{\bm{r},0} \exp( i \bm{q} \cdot \bm{r})$. 
In the standard treatment \cite{Moriya-Taki}, Eq.~(\ref{eq:DS}) 
is expanded around an antiferromagnetic 
wave-vector $\bm{Q}$, which is appropriate and has been used for 
weak itinerant antiferromagnets of $d$-electron systems 
\cite{MoriyaBook85,Moriya-Ueda00}. 
In stead of this expansion, we directly apply Eq.~(\ref{eq:DS}) 
for the present analysis. 
The necessity of the nonexpansion form was suggested 
for heavy fermions because of much weaker $\bm{q}$ dependence 
\cite{Kambe96,Raymond97}; e.g. CeRu$_2$Si$_2$ 
has three antiferromagnetic wave-vectors 
\cite{Regnault88,JRM88,SatoMC99,SatoMC99thesis}. 
To test the SCR theory, hence, we firstly measured 
$\text{Im}[\chi(\bm{q}, E)]$
in a wide $\bm{q}$- and $E$-range at a low temperature and 
fitted the data using Eq.~(\ref{eq:DS}) with the adjustable parameters. 
Secondly we compared the temperature dependence of observed 
$\text{Im}[\chi(\bm{q}, E)]$ and $C/T$ with theoretical predictions. 

A number of constant-$Q$ and -$E$ scans covering an irreducible 
Brillouin zone were performed at $T = 1.5$ K. 
The intensity data were fitted using Eq.~(\ref{eq:DS}), i.e., to 
\begin{subequations}
\label{eq:ImChi}
\begin{eqnarray}
\text{Im}[\chi(\bm{q}, E)] 
&=& 
\chi(\bm{q}) \Gamma(\bm{q})
\frac{ E }{ E^2 + \Gamma(\bm{q})^2 }
\; ,\label{eq:ImChi1}
\\
\chi(\bm{q}) 
&=& 
[\chi_{\text{L}}^{-1} - J(\bm{q})]^{-1} 
\; ,\label{eq:ImChi2}
\\
\Gamma(\bm{q}) 
&=& 
\chi_{\text{L}} \Gamma_{\text{L}}/\chi(\bm{q})
\; ,\label{eq:ImChi3}
\end{eqnarray}
\end{subequations}
with the adjustable parameters $\chi_{\text{L}}$, $\Gamma_{\text{L}}$, 
and $J_{\bm{r},0}$ \cite{para3}. 
Figure~\ref{fig:3Dmap}(a) shows five constant-$Q$ scans selected 
from similar 77 scans. 
One can see from this figure that the fitted curves of these scans well 
reproduce the experiments \cite{single-L}, considering that 
$\chi(\bm{q})$ and $\Gamma(\bm{q})$ of the single Lorentzian, 
Eq.~(\ref{eq:ImChi1}), are determined globally in $\bm{q}$ space 
by the parameters. 

Figures~\ref{fig:3Dmap}(b) and (c) show observed and calculated intensity 
maps of constant-$E$ scans with $E = 1$ meV, viewing the surface of the 
irreducible Brillouin zone. 
These figures show a good quality of fitting, where we note that 
there are 663 observed data points in the irreducible zone. 
The obtained parameters by the fit of the constant-$Q$ and -$E$ scans 
are $\chi_{\text{L}}=0.066 \pm 0.006$ (emu/mol Ce), 
$\Gamma_{\text{L}} \simeq 4.4$ meV $= 51$ K \cite{para3}, and 
fourteen exchange constants listed in Table~\ref{tab:Exc}. 
The number of exchange constants, mathematically equivalent to that of 
Fourier components, were chosen so as to reproduce the three major 
antiferromagnetic correlations peaked at $\bm{k}_{1}$, $\bm{k}_{2}$ 
\cite{Regnault88,JRM88}, and 
$\bm{k}_{3}$ \cite{SatoMC99,SatoMC99thesis} 
and minor structures at the Z and N points, 
but not to show fine structures due to experimental errors.
The requirement of more than ten exchange constants with both 
positive (ferromagnetic) and negative (antiferromagnetic) 
signs reflects the long-range nature of the RKKY interaction. 
To our knowledge, this is the first experimental determination 
of the exchange constants in a paramagnetic heavy-fermion system, 
which has been prohibited for lack of a systematic method. 
\begin{table}
\caption{
\label{tab:Exc}
Exchange constants $J_{\bm{r},0}$ between magnetic moments at 
$\bm{r} = x \bm{a} + y \bm{b} + z \bm{c} $ and $0$. 
They are defined by Eq.~(\ref{eq:DS}) and determined by 
the fit of constant-$Q$ and -$E$ scans shown in Fig.~\ref{fig:3Dmap} 
to the SCR theory [cf. Eqs.~(\ref{eq:ImChi})]. 
A positive constant $J_{\bm{r},0}$ represents a ferromagnetic 
coupling with an exchange energy 
$ - J_{\bm{r},0} \sigma_{\bm{r}} \sigma_{0} $
between Ising variables ($\sigma_{\bm{r}} = \pm 1$) \cite{para3}. 
Errors of exchange constants are $\pm 0.02$. 
}
\begin{ruledtabular}
\begin{tabular}{cccdcccd}
  $x$ & $y$ & $z$ & \multicolumn{1}{c}{$J_{\bm{r},0}$ (K)} 
& $x$ & $y$ & $z$ & \multicolumn{1}{c}{$J_{\bm{r},0}$ (K)} \\
\hline
0 & 0 & 2 & -0.90 & 3 & 0 & 0 &  0.22 \\
1 & 0 & 0 &  0.73 & 0 & 0 & 3 &  0.21 \\
0 & 0 & 1 &  0.66 & 2 & 1 & 1 & -0.12 \\
3 & 0 & 1 &  0.56 & 1/2 & 1/2 & 1/2 &  1.18 \\
2 & 0 & 1 & -0.54 & 3/2 & 1/2 & 1/2 & -0.58 \\
2 & 0 & 0 &  0.47 & 3/2 & 1/2 & 3/2 & -0.39 \\
1 & 0 & 1 & -0.42 & 3/2 & 3/2 & 3/2 & -0.24 \\
\end{tabular}
\end{ruledtabular}
\end{table}

Since we have shown that the phenomenological $\text{Im}[\chi (\bm{q}, E)]$ 
[Eqs.~(\ref{eq:ImChi})] excellently describes the observation at $T=1.5$ K, 
we proceed to the second step at higher temperatures: to compare 
observations with the SCR theory. 
As discussed in Ref.~\cite{Moriya-Taki}, the most important temperature 
dependence of $\chi (\bm{q}, E)$ originates from that of one parameter 
$\chi_{\text{L}}(T)$, and the temperature dependence of the other 
parameters can be neglected in a low temperature range of 
$k_{\text{B}} T \ll \Gamma_{\text{L}}$. 
Under this assumption the $T$ dependence of $\chi_{\text{L}}(T)$ 
can be evaluated by solving 
\begin{equation}
\label{eq:SR}
\sum_{\bm{q}} 
\int_{0}^{E_{\text{c}}}
\frac{\text{d} E}{\pi}
\left[ 1+ \frac{2}{\exp(\beta E)-1} \right]
\text{Im}[\chi(\bm{q},E)] 
= 
\mu^{2}
, 
\end{equation}
where $E_{\text{c}}$ and $\mu$ are the cutoff energy 
and effective moment, respectively. 
This is a sum rule simply implying that the total magnetic scattering crosssection 
integrated in $\bm{q}$ and $E$ spaces is a constant determined by 
the magnetic moment of the crystal-field ground-doublet. 
\begin{figure}
\begin{center}
\includegraphics[width=8.7cm,clip]{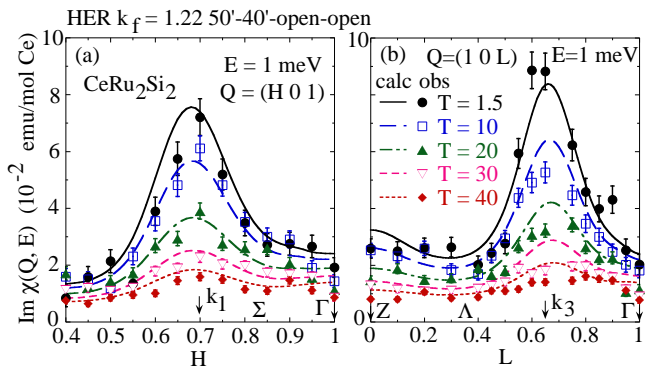}
\end{center}
\caption{
\label{fig:Qscan}
(Color online) Temperature dependence of constant-$E$ scans taken with 
$E = 1$ meV along the $\Sigma$ line $\bm{Q} = (H 0 1)$ (a) 
and the $\Lambda$ line $(1 0 L)$ (b). 
Curves are calculated by the SCR theory using the fit result 
at $T = 1.5$ K and the temperature dependent $\chi_{\text{L}}(T)$.
}
\end{figure}
\begin{figure}
\begin{center}
\includegraphics[width=8.7cm,clip]{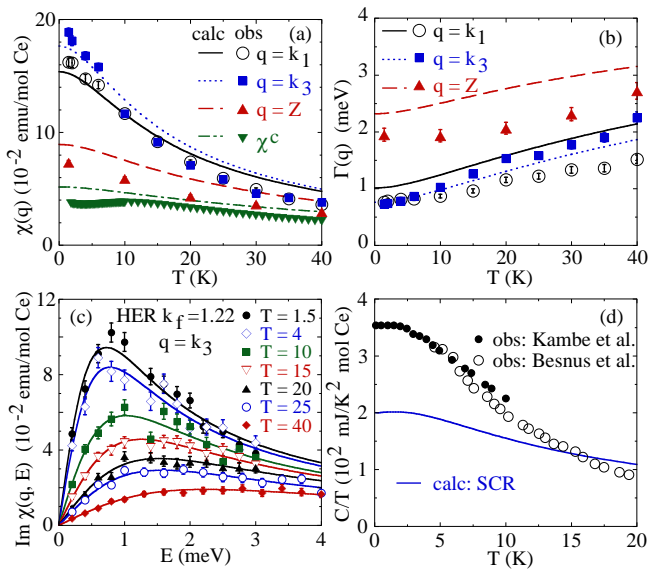}
\end{center}
\caption{
\label{fig:ChiGamCovT}
(Color online) Temperature dependence of the wave-vector dependent 
susceptibility $\chi(\bm{q})$ and uniform susceptibility $\chi^c$ 
(from Ref.~\cite{Haen87}) (a), 
and the energy scale $\Gamma(\bm{q})$ 
at $\bm{q} = \bm{k}_{1}$, $\bm{k}_{3}$ and the Z point (b). 
Curves are the calculation of the SCR theory using the fit result 
at $T = 1.5$ K and the temperature dependent $\chi_{\text{L}}(T)$. 
Observed $\chi(\bm{q})$ and $\Gamma(\bm{q})$ are 
determined by fitting of constant-$Q$ scans to the single Lorentzian form, 
Eq.~(\ref{eq:ImChi1}), at each $\bm{q}$ and $T$. 
These constant-$Q$ scans and fits are shown in (c) for 
$\bm{q} = \bm{k}_{3}$. 
In (d) observed specific-heat coefficient $C/T$ 
(from Refs.~\cite{Besnus85,Kambe96}) is compared with 
the calculation by the SCR theory. 
}
\end{figure}

Several constant-$E$ and -$Q$ scans were carried out at higher temperatures. 
In Fig.~\ref{fig:Qscan} we show constant-$E$ scans along the 
$\Sigma$ and $\Lambda$ lines (see Fig.~\ref{fig:3Dmap}). 
Curves predicted by the SCR theory, i.e., Eqs.~(\ref{eq:ImChi}) 
with $\chi_{\text{L}}(T)$, show agreement with the observation 
within 30 \% in a temperature range $T<40$ K. 
The constant-$Q$ scans were fitted to the single Lorentzian 
[Eq.~(\ref{eq:ImChi1})] with $\chi(\bm{q})$ and $\Gamma(\bm{q})$ 
determined at each $\bm{q}$ and $T$. 
Examples of this fitting are shown in Fig.~\ref{fig:ChiGamCovT}(c) 
for $\bm{q}=\bm{k}_{3}$. 
The resulting $\chi(\bm{q})$ and $\Gamma(\bm{q})$ measured at 
$\bm{q} = \bm{k}_{1}$, $\bm{k}_{3}$, and the Z point are plotted 
as a function of temperature in Figs.~\ref{fig:ChiGamCovT}(a) and (b), 
where the uniform susceptibility is also shown. 
The predicted curves in these figures are in agreement with the 
observations within 30 \% up to $T=40$ K. 
From these comparisons shown in Figs.~\ref{fig:Qscan} and \ref{fig:ChiGamCovT}, 
although there are systematic discrepancies to a certain extent, 
we conclude that the overall agreement of the temperature dependence of 
the observed $\text{Im}[\chi(\bm{q},E)]$ with the SCR predictions is fairly good 
in view of the simple assumption. 

The specific heat is dominated by spin fluctuations at low temperatures, 
and has been used as an important quantity in applying the SCR theory 
\cite{Moriya-Taki,Kambe97}. 
Hence it is interesting to compare the observed specific heat 
\cite{Besnus85,Kambe96} with the SCR theory \cite{spheat}.
%, which can be calculated from observed $\chi(\bm{q},E)$ \cite{spheat}. 
The calculated $C/T$ is compared with the observations in 
Fig.~\ref{fig:ChiGamCovT}(d), which shows agreement within 50 \%.
Considering that the theory employs the same method as that for 
$d$-electron systems, which is not strictly justified for $f$ 
electrons, the agreement is fairly good. 
However, the SCR theory on $C/T$ has to be improved for more 
quantitative purposes, which should agree with a FL theory of 
renormalized quasiparticles \cite{Zwicknagl92} at $T=0$. 

We have shown that the SCR theory \cite{Moriya-Taki} provides a useful 
description of the spin fluctuations, i.e., $\chi(\bm{q},E)$, for 
the archetypal heavy fermion CeRu$_2$Si$_2$. 
It seems that this phenomenology will be applicable to a certain class 
of heavy-fermions. 
The success for CeRu$_2$Si$_2$ is partly based on the fact that 
Eq.~(\ref{eq:DS}) is a good approximation at low temperatures. 
In fact, almost the same form of equation was derived by 
a $1/d$-expansion theory \cite{Ohkawa98} on the periodic Anderson model. 
At high temperatures the SCR theory proposed the simple 
assumption that only one parameter $\chi_{\text{L}}(T)$ depends 
on temperature, and used the exact sum rule of Eq.~(\ref{eq:SR}). 
In the limit $T \rightarrow \infty$, this recipe gives the 
correct Curie-Weiss susceptibility 
$\chi(\bm{q}) = \mu^{2}/[k_{\text{B}}T - J(\bm{q})]$. 
This natural extension may account for why we obtain the 
acceptable agreement up to 40 K (see Figs.~\ref{fig:Qscan} 
and \ref{fig:ChiGamCovT}) beyond the original expectation, 
$T \ll \Gamma_{\text{L}}/k_{\text{B}} \simeq 51$ K \cite{para3}. 

The three antiferromagnetic wave-vectors $\bm{k}_1$, $\bm{k}_2$, 
and $\bm{k}_3$ are determined by 
$J(\bm{q})=$ max [cf. Eq.~(\ref{eq:DS})]. 
In the SCR theory $J(\bm{q})$ is ascribed mainly to the RKKY mechanism 
and is weakly temperature dependent. 
However one also might attribute the wave-vectors to nestings of the 
quasiparticle bands. 
According to Ref.~\cite{Ohkawa98}, the nesting mechanism can be 
incorporated in Eq.~(\ref{eq:DS}) by replacing $J(\bm{q})$ with 
$J_{s}(\bm{q})+J_{Q}(\bm{q},E)$, where the first term represents 
exchange interactions involving intermediate states of 
high-energy excitations (including the RKKY interaction), and the 
second term involving low-energy excitations of the quasiparticles. 
This second term possesses the energy scale $k_{\text{B}} T_{\text{K}}$, 
and consequently would show larger $T$ dependence \cite{Ohkawa98}. 
We speculate that $J_{Q}(\bm{q},E)$ is smaller 
than $J_{s}(\bm{q})$ for CeRu$_2$Si$_2$ (in zero magnetic field), 
because nesting wave-vectors close to 
$\bm{k}_1$, $\bm{k}_2$, and $\bm{k}_3$ cannot be easily seen in 
the Fermi surfaces of the band structure \cite{Zwicknagl92}, 
and because $T$ dependence of $J(\bm{q})$ could be neglected in our 
analysis. 
On the other hand, the quasiparticle mechanism supposedly brings 
about ferromagnetic exchange interactions under magnetic fields 
close to the metamagnetic crossover \cite{Sato-Ohkawa01}. 
Therefore, to resolve these problems on the quasiparticle 
contribution, improved SCR theories based on microscopic models 
\cite{Saso99-Pepin99} are awaited. 
It will be also interesting to apply them to those 
in which the quasiparticles play essential roles in 
spin fluctuations, e.g. UPt$_3$ \cite{Aeppli88} 
and CeNi$_2$Ge$_2$ \cite{Kadowaki03}. 

Singularities of QCPs in itinerant magnets have been 
thought to be described by the SCR theory 
\cite{MoriyaBook85,Moriya-Ueda00} or equivalently 
by the spin-density-wave type fixed point \cite{Hertz76-Millis93}. 
However a recent study of criticality of a heavy-fermion system 
CeCu$_{6-x}$Au$_x$ tuned to a QCP \cite{Schroder00} caused 
controversy on a possibility of a locally-critical fixed point 
\cite{Coleman01,Si01}. 
A key observation, supporting this fixed point,
is the $E/T$ scaling form 
$\text{Im}[\chi(\bm{Q},E)] =T^{-\alpha}g(E/T)$ 
at the antiferromagneic wave vector $\bm{Q}$ \cite{Schroder00}. 
In contrast, the SCR theory predicts the $E/T^{3/2}$ scaling form 
$\text{Im}[\chi(\bm{Q},E)] = T^{-3/2}g(E/T^{3/2})$, 
which one can see using $\chi(Q)^{-1} \propto T^{3/2}$ at the QCP 
\cite{Moriya-Ueda00} and Eqs.~(\ref{eq:ImChi}). 
For the present SCR theory of CeRu$_2$Si$_2$, located slightly off 
the QCP, this characteristic $T^{3/2}$ dependence appears 
approximately as $\Gamma(\bm{Q}=\bm{k}_i) \simeq A + B T^{3/2}$ 
in a low temperature range $0<T<10$ K. 
Hence we speculate that it may be interesting to accurately measure 
$T$ dependence of $\Gamma(\bm{Q})$ to determine  whether it shows 
the SCR $T^{3/2}$-dependence or is closer to $T^{1}$-dependence, which 
would suggest the $E/T$ scaling. 

In conclusion, we have demonstrated that the SCR theory remarkably 
well describes the spin fluctuations of the archetypal heavy-fermion 
CeRu$_2$Si$_2$. 
The analysis using magnetic excitation data covering the entire 
irreducible Brillouin zone has enabled us to determine the 
fourteen exchange constants. 

We wish to acknowledge B. F{\aa}k, J. Flouquet, F. J. Ohkawa, S. Raymond, 
and Y. Tabata for discussions.

\bibliography{ceru2si2}

\begin{thebibliography}{31}
\bibitem[{par()}]{para3}
\bibinfo{note}{Since Eq.~(\ref{eq:ImChi3}) does not account for an experimental
  fact that $\Gamma(\bm{q}) \sim 0$ close to $q \sim 0$, we modify it to ad hoc
  $\Gamma(\bm{q})^{-1} =\chi(\bm{q}) \{
  c_{1}+c_{2}/[(ha^{*})^2+(kb^{*})^2+c_{3}(lc^{*})^2]^{1/2} \}$. This is an
  interpolation between Eq.~(\ref{eq:ImChi3}) for large $q$ and $\Gamma(\bm{q})
  \propto q$ for small $q$, which is neglected in Ref.~\cite{Moriya-Taki} for
  antiferromagnetic cases. Hence the actual fit parameters are
  $\chi_{\text{L}}$, $c_{i}$ ($i=1,2,3$), and $J_{\bm{r},0}$. The values of
  $c_{i}$ obtained by the fit at $T=1.5$ K are $c_{1}=3.4 \pm 0.2$
  [meV~emu/(mol~Ce)]$^{-1}$, $c_{2}=1.5 \pm 0.1$ [meV~\AA~emu/(mol~Ce)]$^{-1}$,
  and $c_{3}=2.8 \pm 0.4$. The energy scale $\Gamma_{\text{L}}$ is approximated
  as $\Gamma_{\text{L}} \simeq 1/(c_{1} \chi_{\text{L}}) = 4.4$ meV. When
  $J_{\bm{r},0}$ in Table~\ref{tab:Exc} is used in Eq.~(\ref{eq:ImChi2}), a
  conversion factor $0.58$ [K~emu/(mol~Ce)]$^{-1}$ is multiplied.}
\bibitem[{sin()}]{single-L}
\bibinfo{note}{The single Lorentzian [cf. Eq.~(\ref{eq:ImChi1})] is the
  simplest approximation for $\text{Im}[\chi(\bm{q}, E)]$ in a low energy
  range. We think the quality of fitting using the single Lorentzian shown in
  Fig.~\ref{fig:3Dmap}(a) is good enough for the present purpose. It should be
  noted that the double Lorentzian form for $\text{Im}[\chi(\bm{q}, E)]$ used
  in Ref.~\cite{JRM88} has been somewhat overemphasized. In fact, the single
  Lorentzian provides good quality of fitting
  [Refs.~\cite{Raymond97,Raymond98}; see Fig.~\ref{fig:ChiGamCovT}(c)].}
\bibitem[{sph()}]{spheat}
\bibinfo{note}{We used Eq.~(4.2) in Ref.~\cite{Moriya-Taki}, where the factor 3
  was replaced by 1, accounting for the strong uniaxial anisotropy.}
\end{thebibliography}

\end{document}